# An Interpretable Transformer-Based Foundation Model for Cross-Procedural Skill Assessment Using Raw fNIRS Signals

Subedi A., De S., Cavuoto L., Schwaitzberg S., Hackett M., Norfleet J.

Abstract:

Objective skill assessment in high-stakes procedural environments requires models that not only decode underlying cognitive and motor processes but also generalize across tasks, individuals, and experimental contexts. While prior work has demonstrated the potential of functional near-infrared spectroscopy (fNIRS) for evaluating cognitive-motor performance, existing approaches are often task-specific, rely on extensive preprocessing, and lack robustness to new procedures or conditions. Here, we introduce an interpretable transformer-based foundation model trained on minimally processed fNIRS signals for cross-procedural skill assessment. Pretrained using self-supervised learning on data from laparoscopic surgical tasks and endotracheal intubation (ETI), the model achieves >88% classification accuracy on all tasks, with Matthews Correlation Coefficient exceeding 0.91 on ETI. It generalizes to a novel emergency airway procedure—cricothyrotomy—using fewer than 30 labeled samples and a lightweight (<2k parameter) adapter module, attaining an AUC greater than 87%. Interpretability is achieved via a novel channel attention mechanism—developed specifically for fNIRS—that identifies functionally coherent prefrontal sub-networks validated through ablation studies. Temporal attention patterns align with task-critical phases and capture stress-induced changes in neural variability, offering insight into dynamic cognitive states.

1. Introduction

The objective, scalable assessment of procedural expertise—particularly in high-stakes domains such as surgery and emergency medicine—remains an unsolved problem [Levin et al., 2019; Makary & Daniel, 2016]. Current methods largely rely on subjective observation and rating, which introduce variability and bias, lack physiological grounding, and do not generalize across tasks or individuals. While functional near-infrared spectroscopy (fNIRS) has emerged as a promising non-invasive tool for measuring neural activity during task performance [Nemani et al., 2018; Gao et al., 2021; Keles et al., 2021], prior models—including our own—have largely relied on extensive preprocessing pipelines, lacked physiological interpretability, and demonstrated limited generalization across tasks or stress conditions. Recent sub-region-based analyses have shown that localized activation patterns in prefrontal and motor-related areas can distinguish between proficient and non-proficient performers [Nemani et al., 2018, Gao et al., 2021, Subedi et al., 2025]. The complex spatiotemporal dynamics captured in these signals contain rich information about cognitive and motor processes that could enable more robust, generalizable assessments across diverse settings. However, existing models tend to be task-specific and require retraining when applied to new procedures, impeding scalability.

These limitations highlight the need for a domain-specific foundation model—an architecture trained to extract reusable neural representations across various tasks, individuals, and conditions.

In contrast to conventional pipelines, such a model would support modular adaptation to new contexts using minimal data, while retaining interpretability. Foundation models, initially developed in natural language processing and vision [Lu et. al., 2019; Radford et al., 2021; Dosovitskiy et al., 2021], are pretrained on broad datasets and fine-tuned efficiently for downstream tasks. We extend this concept to neuroimaging by proposing a transformer-based foundation model trained directly on raw fNIRS time-series signals.

Advancing neuroimaging-based skill assessment requires addressing two critical scientific challenges. First, the simultaneous relationship between specific brain regions, their temporal activation patterns, and skill proficiency often remains opaque in existing models, limiting insights into underlying neurophysiological mechanisms [Andersen et al., 2024]. Without interpretable spatiotemporal mapping of neural correlates during task execution, targeted interventions for skill development remain difficult to implement. Second, the remarkable variability in neural responses across tasks, individuals, and experimental contexts necessitates frameworks capable of efficient adaptation without complete retraining [Parisi et al., 2019].

Machine learning (ML) and deep learning (DL) approaches applied to fNIRS data have achieved high classification performance in skill assessment [Nemani et al., 2018; Nemani et al., 2019; Gao et al., 2021; Keles et al., 2021; Wang et al., 2022]. However, these studies typically require extensive multi-stage signal preprocessing before model training, which is time-consuming, potentially removes subtle yet informative signal features, and impedes real-time deployment [Eastmond et al., 2022; Yücel et al., 2017]. Our prior work established that raw fNIRS signals contain sufficient information for high-accuracy skill classification via end-to-end deep learning [Subedi et al., 2025], suggesting that minimally processed neuroimaging data could support both advanced interpretability methods and flexible adaptation approaches previously unexplored in this domain.

Standard ML/DL models also suffer from sensitivity to distribution shifts when applied to out-of-distribution data [Quinonero-Candela et al., 2009; Arjovsky et al., 2019; Liu et al., 2021], requiring substantial retraining or risking catastrophic forgetting [Parisi et al., 2019]. While parameter-efficient fine-tuning techniques have addressed these challenges in language models [Houlsby et al., 2019; Pfeiffer et al., 2020; Han et al., 2024], such strategies have yet to be systematically applied in the context of neuroimaging.

Additionally, standard Explainable AI (XAI) methods, such as Gradient-weighted Class Activation Mapping (Grad-CAM), developed for spatial CNN analysis [Selvaraju et al., 2017], face challenges when applied to variable-length, potentially noisy fNIRS time series. Issues include coarse temporal localization and adapting gradient-based methods effectively to capture complex sequential dynamics [Schlegel et al., 2022], motivating the exploration of architectures inherently better suited for interpretable sequence analysis. While XAI for time-series is an emerging field [Farahani et al., 2022], existing studies predominantly apply post-hoc methods like SHAP to models trained on conventionally processed data [Shibu, 2023] or explore transformer architectures primarily for classification gains rather than deep, integrated interpretability and efficient adaptability from raw signals [Wang et al., 2022]. **Thus, a critical gap remains in** developing end-to-end interpretable frameworks that operate directly on raw fNIRS data and possess demonstrated capabilities for efficient generalization and adaptation.

To address these limitations, we propose an end-to-end deep neural network framework based on a transformer architecture. This framework is designed to: (a) pretrain on raw fNIRS time-series via self-supervised masked segment modeling to produce a reusable, domain-specific foundation model [Devlin et al., 2019; Kostas et al., 2021], (b) provide integrated, interpretable spatiotemporal explanations of motor skill classification and cognitive state using novel, dedicated channel and temporal attention mechanisms; and (c) enable parameter-efficient adaptability to new procedures via lightweight adapter modules [Houlsby et al., 2019; Han et al., 2024; Wan et al., 2024]. The interpretability of the framework is evidenced by the functional relevance of identified channels through ablation studies and alignment with established functional neuroanatomy, while correlating temporal attention dynamics with critical task phases across performance and stress conditions.

Figure 1 provides an overview of our proposed end-to-end framework and validation workflow. We leverage fNIRS collected across three procedural contexts—laparoscopic suturing, endotracheal intubation (ETI) under both baseline and cognitively loaded (stress) conditions—for model pretraining and supervised training, and use a fourth dataset involving cricothyrotomy, a distinct emergency airway procedure, to evaluate generalization (Fig. 1a, 1b; table inset).

Our approach consists of three key stages. First, a transformer encoder backbone is pre-trained on raw fNIRS signals using a self-supervised masked segment reconstruction objective, enabling the model to learn spatiotemporal features without explicit labels (Fig. 1c). Second, this pretrained encoder (with frozen weights) is paired with a decoder that includes both channel and temporal attention modules and is trained on downstream classification tasks (pass/fail) using labeled examples from the same procedures. This stage yields both predictive performance and interpretable attention maps (Fig. 1d). Held-out validation is performed on left-out trials from the same training set (4th pass repetition for each subject). Third, to assess generalizability, we evaluate parameter-efficient adaptation to different procedure—cricothyrotomy—by inserting a lightweight (<2k parameters) adapter module while keeping both the encoder and decoder frozen (Fig. 1e). The model is trained with limited examples (e.g., 30-shot adaptation) and tested on unseen subjects. The fidelity of the XAI mechanisms is further validated through targeted channel ablation studies and an analysis of temporal attention alignment with defined procedural phases and stress-induced neural variability (see Fig. 5).

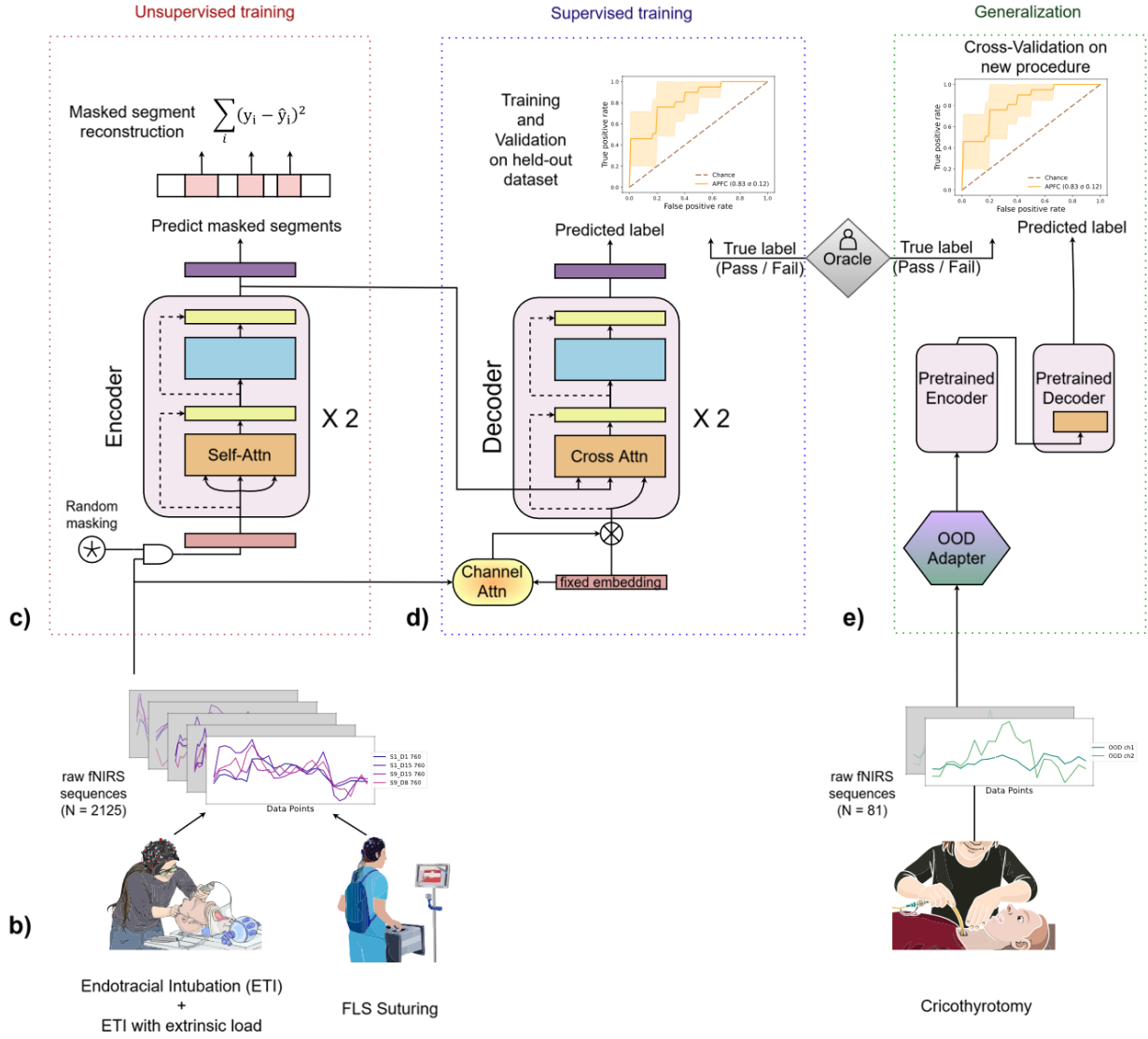

Figure 1. **Overview of the proposed transformer-based foundation model and cross-procedural generalization pipeline**. a) fNIRS datasets from four study arms spanning three procedures – laparoscopic suturing, endotracheal intubation (ETI) with and without cognitive load, and cricothyrotomy. b) Illustrative depictions of the procedural tasks. c) Self-supervised masked segment reconstruction is used to pretrain the encoder on raw fNIRS signals, capturing spatiotemporal dynamics across the prefrontal cortex. d) In the supervised phase, the pretrained encoder is paired with a decoder incorporating channel and temporal attention modules to classify task outcomes (success/failure). **e)** A lightweight adapter module enables parameter-efficient generalization to a new task (cricothyrotomy) using a small number of labeled examples.

## 2. Results

We evaluated the performance of our transformer-based foundation model across three core objectives: (1) accurate classification of procedural outcomes using raw fNIRS signals, (2) interpretability of the model's learned representations, and (3) generalization to an out-of-distribution (OOD) surgical task using parameter-efficient adaptation. We report performance on supervised classification of FLS Suturing and Endotracheal Intubation (ETI) under normal and stress conditions, followed by evaluation on a novel cricothyrotomy task with limited training data.

The masked pretraining stage converged to a mean-squared-error value of 0.053, and training is terminated when improvements fell below a threshold (1e-3) for a preset number of epochs (300). Nine different models were trained with different initialization properties to reduce variance in reported scores and attention-based interpretations. The decoder was trained for each task separately, since the annotations follow distinct evaluation rubrics. Performance was always measured in the held-out set. The results from all nine pretrained encoders are averaged and are reported in Fig 2.

Our deep learning model achieves accuracies of over 88% across all the tasks (Fig 2a). For both study arms involving the ETI task, it achieves an MCC greater than 95%. For FLS suturing task, which is a longer procedure, it achieves an MCC of 64.6%. With a sensitivity of 90.9%, the model correctly identifies 90.9% of the failed surgical trials. Complementarily, it achieves a specificity of 76.9%, correctly identifying 76.9% of the successful trials. These scores are substantially higher for the ETI task, indicating stronger discrimination between successful and unsuccessful intubation attempts.

The supervised-training achieves an average area under the curve (AUC) the Receiver Operating Characteristics (ROC) curve of 0.92, 0.99 and 0.92 for FLS Suturing, ETI, and ETI with cognitive load, respectively (Fig. 2b). AUCs vary by no more than 0.01 across all pretrained encoders at 95% confidence intervals. Similar Precision-Recall (PR) analysis (Figure 2c) yields AUCs greater than 0.96, with the lower bound of confidence interval exceeding 0,985 for all the tasks (see "Methods", Statistical Analysis). The consistently high PR-AUC values indicate that the model maintains discriminative power even under class imbalance—a common challenge in skill assessment scenarios.

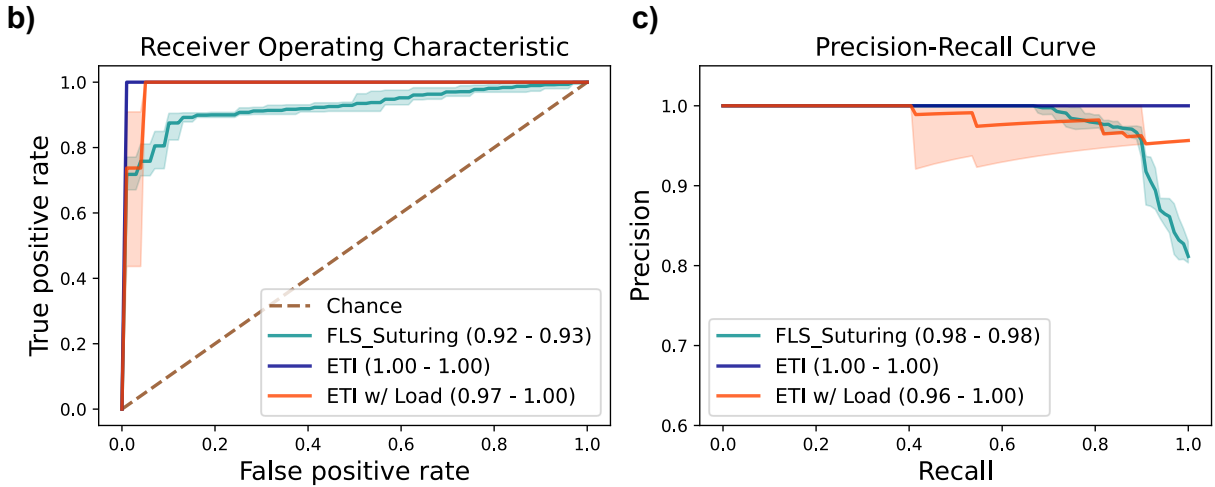

Figure 2. **Validation results for three procedural tasks**. **a)** Classification metrics for each task. **b)** Receiver Operating Characteristic (ROC) curves and **c)** Precision-Recall (PR) curves, each shown with 95% confidence intervals.

The model was further validated on an Out-of-Distribution dataset. The adapter layer is used to transform the brain activations from the new task to match the distributions of the pretrained model. The OOD dataset differs both in terms of the procedural context and optode montage from the training dataset. The adapter module (<2k parameters) is used to adapt the to this new setting with only 81 labeled samples. Leave one subject-out cross validation yields a mean accuracy of 87.7%. (Figure 3b), demonstrating that the model can accurately classify success or failure in cricothyrotomy trials from previously unseen subjects. Thus, on average, it can correctly predict the success/failure of 87% of the trials carried out by a new human subject. This level of generalization—achieved without any updates to the pretrained encoder or decoder—suggests that the model has captured fundamental neural dynamics that are transferable across procedural domains.

To further determine the minimum number of labeled samples required for adaptations, we performed a few-shot learning analysis using multiple few-shot samples 10, 20, and 30 labeled examples. Prediction performance was evaluated on the validation set using 3-fold cross-validation. Figure 3c shows the box plots for accuracy, sensitivity, and specificity, while the corresponding ROC curves are shown in Figure 3d. Results indicate that 30-shot training—with 15 samples per class—is sufficient to yield ROC AUC greater than 0.89, with both sensitivity and specificity exceeding 75%. Performance improves consistently with increasing shot count, confirming the utility of the adapter mechanism for scalable deployment with constrained data availability. While performance is lower than full cross-validation with all available OOD samples, this analysis provides a practical estimate of the sample size needed for effective adaptation to a new procedural task.

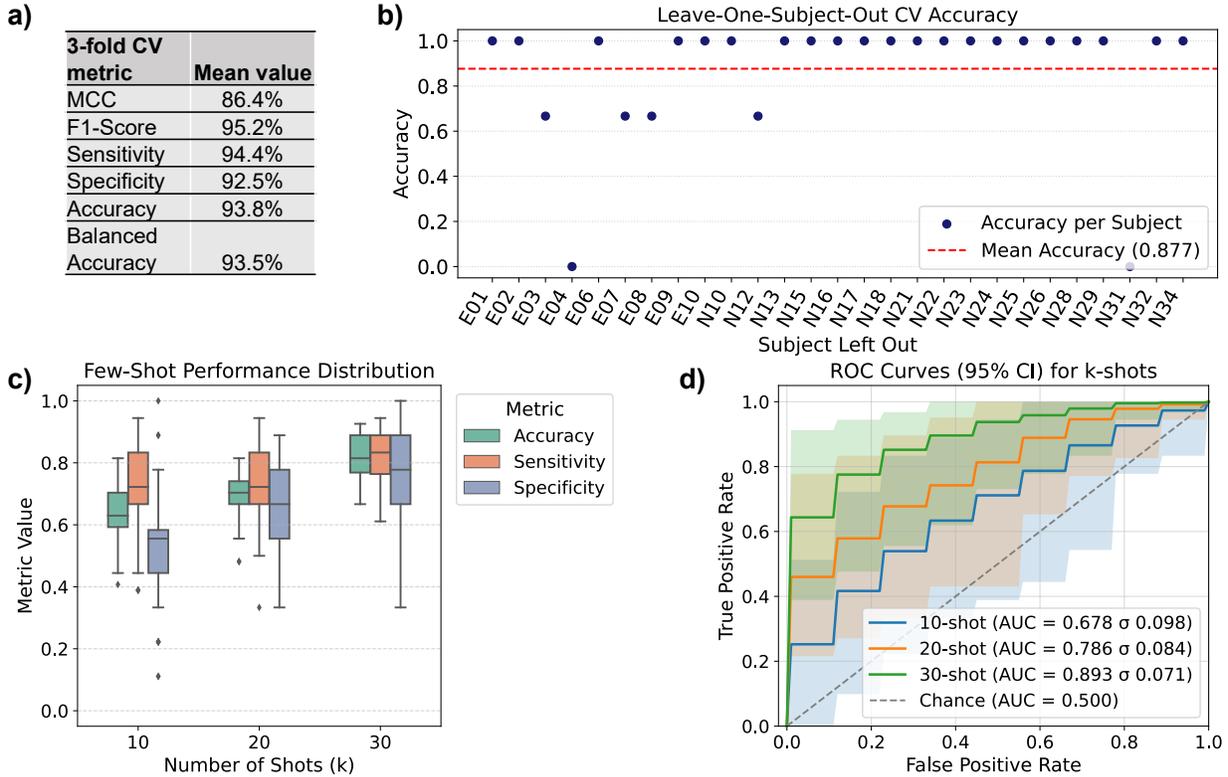

Figure 3. **Experimental results on the out-of-distribution (OOD) procedure (cricothyrotomy).** **a)** Mean classification metrics for 3-fold Cross-Validation **b)** Prediction accuracies from leave-one-subject-out cross validation. **c)** Boxplots of prediction performance metrics of adapter module on the validation set following k-shot adapter training. **d)** Corresponding ROC curves with 95% confidence intervals and standard deviation of AUC

3. Discussion

3.1. Spatial (Channel) Attention

The channel attention mechanism consistently identified a sparse, functionally coherent sub-network within the prefrontal cortex (PFC) montage (Fig 4) that contributes maximally to motor performance classification. This sub-network remained stable across all nine model initializations (see Fig S1), comprising specific loci within the left dorsolateral (DLPFC) and dorsal PFC, alongside bilateral frontopolar cortex (BA10).

These highlighted regions correspond to established nodes of the cognitive control networks that support goal-directed behavior [Cole & Schneider, 2007; Vincent et al., 2008]. The DLPFC and dorsal PFC have been extensively associated with core executive functions, including working memory for task-relevant information and the sequencing of planned actions [Miller & Cohen, 2001; Ridderinkhof et al., 2004; Mandrick et al., 2016]. Complementing these, the bilateral frontopolar cortex sits at the apex of the executive hierarchy, integrating action outcomes, maintaining long-term task goals, and supporting strategic decision making in complex procedural settings such as the ETI [Burgess et al., 2007; Koechlin & Hyafil, 2007; Spiers et al., 2015; Boschin et al., 2015; Mansouri et al., 2020; Kroger et al., 2022].

The coactivation of planning and high-level control areas revealed by our explainable AI (XAI) framework, linking action execution (DLPFC) with strategic integration (frontopolar cortex), reflects the multifaceted cognitive demands imposed by surgical and emergency procedures [Badre & D'Esposito, 2009; Koechlin et al., 2003]. The reproducibility of this attention-based sub-network across trials and conditions strongly suggests that the model's decisions are anchored in meaningful neurophysiological signals, rather than in spurious features or overfitting.

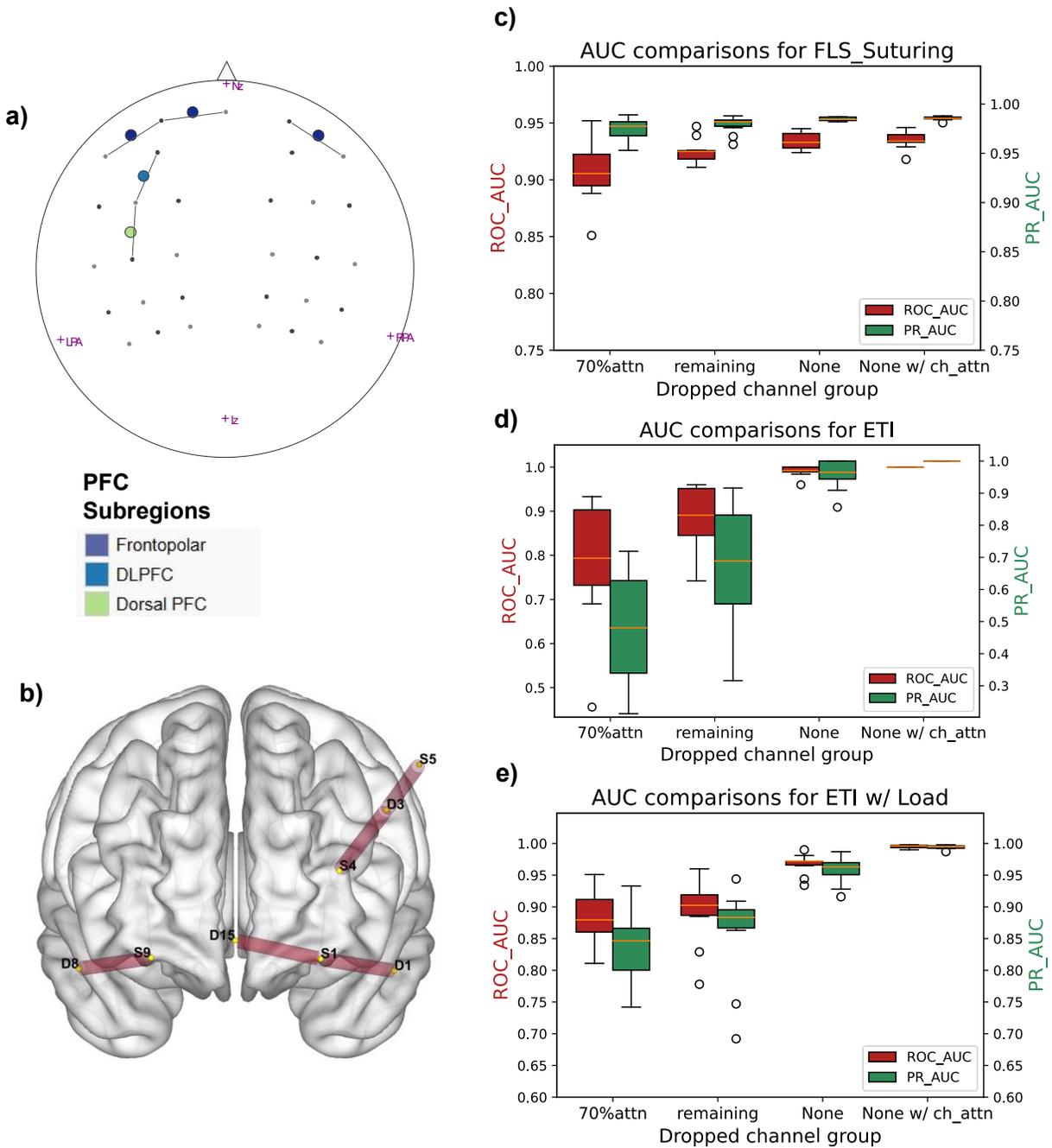

Figure 4. **Spatial interpretability via channel attention and ablation analysis**. **a)** Top-view schematic of PFC montage showing cortical channels accounting for ~70% of attention weights, categorized by subregion (Frontopolar, DLPFC, Dorsal PFC. **b)** Anatomical projection using AtlasViewer showing spatial locations and optode numbers of the high-attention channels. **c)** ROC and PR AUCs following systematic ablation of channel groups for the FLS Suturing task. **d)** AUC comparisons for the ETI task under cognitive load. In each task, performance is reported for four conditions: top 70% attention channels dropped ('70%attn'), remaining channels dropped ('remaining'), all channels retained without attention ('None'), and all channels retained with attention active ('None w/ ch_att').

To empirically validate the functional relevance of the cortical channels identified by the attention mechanism, we conducted an ablation analysis. We systematically assessed the impact on classification performance (measured by ROC AUC and PR AUC, chosen for their cutoff-independence) when removing either the subset of channels receiving the highest attention scores (cumulatively accounting for approximately 70% of total attention weight; '70%attn') or the remaining, lower-attention channels ('remaining'). The results, presented for all three procedures (Figure 4c-e), provide strong quantitative support for the fidelity of the channel XAI.

Across all conditions and metrics, removing the top-attended channels resulted in a demonstrably greater performance detriment (i.e., lower median AUC values and often increased variance) compared to removing the less attended, remaining channels. This effect was especially pronounced in the cognitively loaded ETI task (Fig. 4e), where removing the high-attention channels led to a marked drop in both ROC and PR AUC. In contrast, dropping the remaining channels produced only minor degradation relative to the baseline performance using all channels ('None').

These findings confirm that the channel attention mechanism reliably identifies and prioritizes the cortical channels carrying the most discriminative information pertinent to skill classification tasks. Performance with all channels included ('None') confirms expected baselines, while the highest AUC scores are consistently obtained when the channel attention module is active ('None w/ ch_att'), underscoring the utility of learned spatial weighting for enhancing classification accuracy. This pattern holds across tasks, indicating that attention-driven spatial weighting confers consistent benefit over fixed, unweighted architectures.

Taken together, these ablation results provide compelling empirical support for interpretability and functional validity of the channel attention maps. The empirical 70% attention threshold emerges as a stable and biologically plausible marker of the most informative neural subspace for procedural skill classification.

### 3.2. Temporal Attention

We further analyzed the insights in temporal domain yielded by the built-in attention framework by examining the consistency of neural activation across the most salient cortical channels during distinctly defined ETI task phases [Kunkes et.al, 2022]. Specifically, we computed the standard deviation (STD) of the average HbO signal across the top-performing channels identified by the channel attention mechanism for each subject during the first repetition (Figure 5). Lower STD values across these task-critical channels are interpreted as reflecting greater functional synchronization or more consistent, coordinated network engagement. Reduced neural variability has been linked to more stable task engagement during demanding cognitive processes, such as working memory [Lundqvist et al., 2018; Shine et al., 2016]. Furthermore, a more pronounced 'quenching' of such variability is predictive of superior perceptual performance, suggesting more consistent and efficient neural encoding [Arazi et al., 2017].

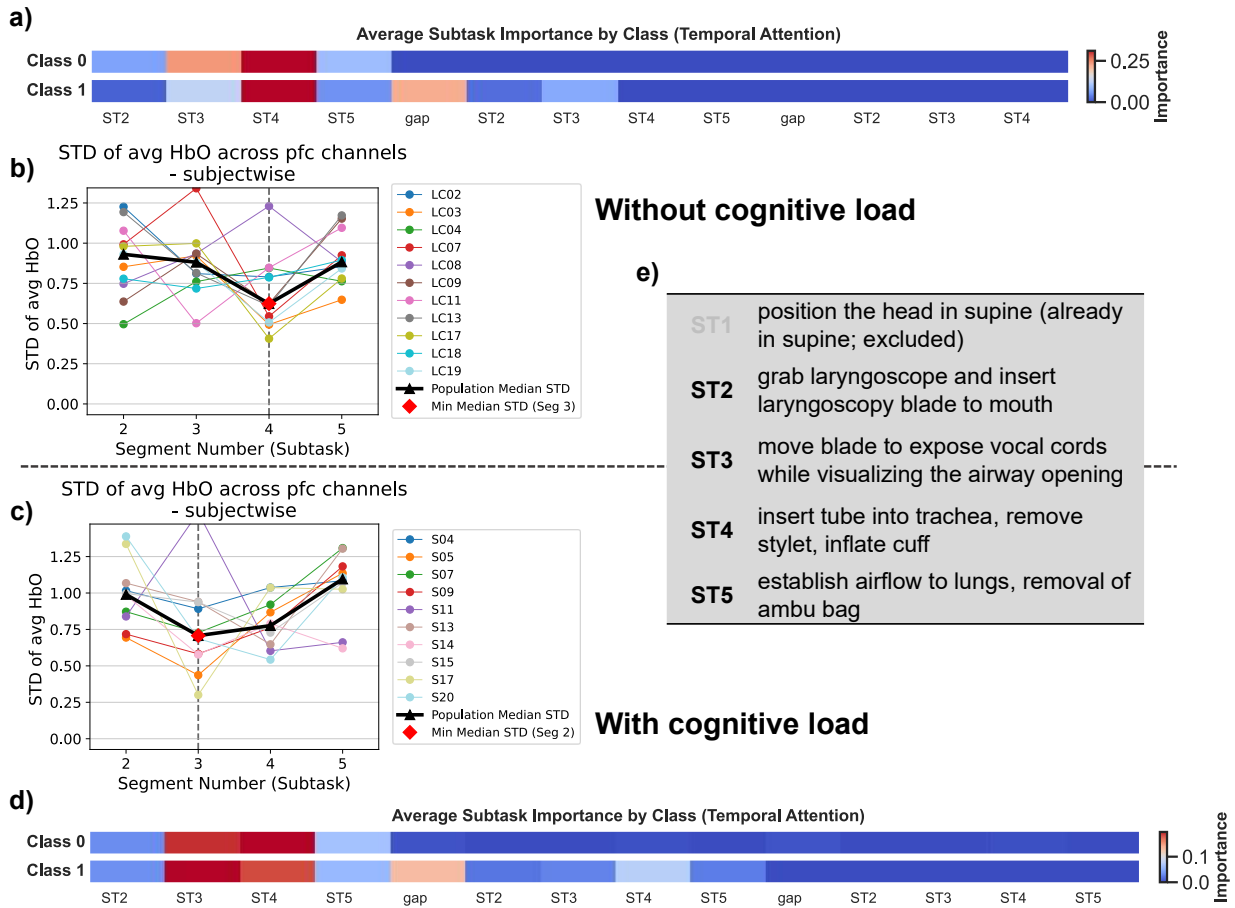

Figure 5. **Insights from the temporal attention module.** Analysis carried out for ETI task without load **(top half)** and with load **(bottom half)** to highlight differential activations. Standard deviations in raw signals averaged over sub-task segments shown for color-coded individual subjects and the population in black under **b)** normal and **c)** stressed conditions. Correspondingly averaged temporal attention scores provided by the model are shown for individual successful and unsuccessful examples are shown in the heatmaps under **a)** normal and **d)** stress. **e)** ETI cognitive task phases as defined in [Kunkes 2022]

Under normal (no stress) conditions (Figure 5a and b), the population median STD of the HbO signal across attention-weighted channels reached its minimum during Subtask 4, the endotracheal tube insertion phase. This suggests that in the absence of external stressors, the identified sub-network achieves peak coordination and consistency during the most demanding procedural step.

Such reduced neural variability has been associated with more stable task engagement and enhanced cognitive performance [Lundqvist et al., 2018; Arazi et al., 2017], and in well-learned visuomotor skills, may even coincide with decreased overall PFC activation [Floyer-Lea & Matthews, 2004]. This period of heightened consistency in neural signals aligns with the temporal attention weights from our model (Figure 1d), which also predominantly highlights Subtask 4, indicating a convergence of efficient neural processing and model-driven attentional focus.

A distinct pattern emerged under stress (Figure 5c and d), where the minimum median STD shifted earlier to Subtask 3 (laryngoscope insertion and initial visualization), and the temporal attention became more distributed across Subtasks 2 and 3. This shift suggests that acute stress reorganizes the timing and duration of coordinated neural responses, with peak synchronization occurring earlier and sustaining longer. Such an anticipatory shift in neural resource deployment is consistent with stress-induced adaptations in large-scale brain networks and changes in PFC function aimed at managing heightened preparatory demands and cognitive load [Hermans et al., 2014; Arnsten, 2009; Shackman et al., 2011; Robinson et al., 2014]. The broadened temporal focus likely reflects prolonged cognitive engagement and greater mental effort, a phenomenon well-documented in studies that associate increased PFC activation with higher cognitive load and mental effort [e.g., Cheng et al., 2024; Mandrick et al., 2016].

Thus, the framework reveals physiologically interpretable changes in both the timing and duration of peak neural activity, combining direct analysis of raw signal variability with model-derived temporal attention. These converging results demonstrate that the XAI framework not only captures task structure but also reflects stress-induced alterations in cognitive control dynamics, providing a window into the brain's adaptive state changes during real-time procedural performance.

### 3.3. Translation

These spatiotemporal neural patterns offer insights that extend beyond direct training feedback. By identifying the critical sub-tasks and their corresponding neural signatures (e.g., minimal STD and focused temporal attention during specific ETI phases), the framework can be applied to dissect other complex procedural skills where key performance actions are less well-defined. This could refine task decomposition and instructional design, particularly in domains where the hierarchy of subtasks remains ambiguous.

Identifying sparse but critical prefrontal networks also has implications for neuro-monitoring hardware design. For instance, these findings could inform the development of streamlined fNIRS headgear with fewer but optimally placed optodes, improving wearability and ecological validity in field conditions [Pinti et al., 2018; Pinti et al., 2019]. Alternatively, the attention-informed spatial maps could guide multimodal neuroimaging configurations, such as hybrid EEG–fNIRS systems, to achieve greater signal specificity and reduce noise [Ahn et al., 2016; von Lühmann et al., 2017; Chiarelli et al., 2017].

Furthermore, the observed neural dynamics—specifically the timing and consistency of PFC network coordination (via STD minima) and the model's temporal attention distributions—can serve as objective inputs for real-time cognitive state monitoring. Deviations from normative attentional or variability signatures, such as the shift from Subtask 4 (no stress) to Subtask 3 (stress), may indicate deteriorating performance or escalating cognitive load in operational environments [Aricò et al., 2017; Dehais et al., 2020].

Finally, understanding these context-dependent changes in neural coordination and attentional focus may enable targeted cognitive interventions. For example, the XAI-derived timing markers could be used to guide closed-loop neuromodulation protocols (e.g., transcranial direct current stimulation) to selectively enhance engagement, working memory, or task-switching capacities [Bikson et al., 2018; Polanía et al., 2018].

While these translational pathways remain exploratory, the framework presented here offers a novel, interpretable toolset for analyzing, monitoring, and potentially enhancing human cognitive performance across a range of complex, real-world tasks.

4. Study Limitations:

Although explainability provides rich and physiologically plausible insights into the relationship between model predictions and the raw fNIRS signals, there are limitations to the interpretability of temporal attention. Specifically, due to the lack of synchronization between the available procedural videos and neuroimaging signals, a trial-by-trial alignment between hemodynamic responses and precise behavioral events is challenging. Future work incorporating curated and manually aligned video data will enable more precise temporal validation, allowing finer-grained comparisons between neural activation patterns and task execution.

In addition, the model's spatial coverage was limited to prefrontal cortex subregions to ensure trainability and model convergence. Attempts to train on the full montage led to convergence issues. While executive and cognitive functions, typically localized to the PFC, have been linked to surgical proficiency in prior work, tasks involving sensorimotor or parietal functions may require region-specific retraining.

Exploring whether representations learned from PFC-centric tasks can generalize to other cortical areas is an intriguing future direction, particularly for whole-head or hybrid neuroimaging systems.

Transformer-based models are also known to scale effectively with increasing dataset size. The current training dataset—comprising three procedures and approximately 2,000 labeled trials—represents an important step, but further expansion with additional procedures and participants is expected to improve generalization performance and reduce few-shot sample requirements.

From a clinical perspective, deviations from the normative temporal attention or neural variability patterns identified in this work may serve as candidate biomarkers for executive dysfunction. For instance, atypical signal variability or attentional timing during ETI could flag cognitive strain or dysregulation. However, the presence of individual variability, such as subjects LC08 and S11 who showed atypical STD patterns without consistent performance failure, underscores the need for caution when extrapolating population-level metrics to individuals.

Overall, while the framework shows strong potential for neurophysiological interpretability and cross-task generalization, future studies should aim to integrate richer behavioral data, extend spatial coverage, and account for inter-individual variability to fully realize its translational utility.

5. Methods:

**5.1 Study design and datasets:**

Human subject studies were conducted to collect brain imaging datasets for assessing bimanual motor skills in two procedural tasks: Fundamentals of Laparoscopic Surgery (FLS) suturing with intracorporeal knot tying and endotracheal intubation (ETI). Fundamentals of Laparoscopic Surgery (FLS), a standardized program to teach and evaluate foundational laparoscopic techniques [Higgins et al., 2023], specifically focusing on the psychomotor skill of suturing with intracorporeal knot tying; and endotracheal intubation (ETI), a critical medical procedure involving the insertion of a tube into the trachea to maintain an open airway or administer substances [Endotracheal Intubation]. Two conditions were evaluated for ETI: a baseline (normal) setting and a stress-induced setting, with cognitive load imposed through extrinsic auditory noise to simulate realistic clinical environments. The cognitive load is imposed through the presence of extrinsic noise simulating realistic scenarios. Study protocols were approved by the University at Buffalo Institutional Review Board (IRB) and the US Army Human Research Protection Office (HRPO). Informed consent was obtained from participants after a briefing on the experimental protocol. Neuronal activations were measured via fNIRS during task performance to identify neural correlates of bimanual motor skill proficiency. The study designs are available in Subedi et al, and also provided in the Supplementary Section (Fig S2). Table 1 summarizes the details of the datasets used.

| Study Arm and Task | Participants and Details | Training Protocol and Retention |
|---|---|---|
| **FLS Suturing: Learning Curve and Retention** *Performance assessed using FLS scoring metrics [Fraser 2003]* | **Training group:** 28 students<br>Mean age: 23.57 years (SD 4.38)<br>11 males, 17 females<br>26 right-handed<br>**Control group**: 27 students<br>Mean age: 21.5 years (SD 2.5)<br>10 males, 17 females<br>24 right-handed | - 15-day training<br>- Reps per session: Days 1–5: 3, Days 6–10: 5, Days 11–15: 7<br>- Rest between attempts: 2 min<br>- Max duration per rep: 10 min |
| **Endotracheal Intubation (ETI)** *Performance assessed based on successful intubation within allotted time* | **Training group:** 20 students<br>**Control group**: 19 students | - 3-day training<br>- Total reps: 10 over 3 days<br>- Rest between attempts: 2 min<br>- Max duration per rep: 3 min |
| **Cricothyrotomy** *Performance assessed using …* | **Training group:** 10 Experts, 18 medical trainees | - 3-day training<br>- Reps per session: 6<br>- Rest between attempts: 1 min<br>- Max duration per rep: n.d. |

FLS and ETI procedures were conducted while the neuroimaging signals were acquired using a montage based on the international 10-10 system. Data were collected using a continuous-wave near-infrared spectrometer (NIRSport 2, NIRx Medical Technologies, LLC, NY, USA) sampling at 5.08625 Hz. The headgear configuration consisted of 16 sources and 15 detectors, with source-detector separation of 3 cm, along with 8 short separation detectors placed 0.8 cm from sources. Dual-wavelength infrared light (760 nm and 850 nm) was used for signal acquisition.

Based on previous findings, preprocessing was minimized: raw light intensity signals were converted to optical density and band-pass filtered between 0.01 to 0.5 Hz. Stimulus markers were embedded in the fNIRS data acquired during sessions, which are used to trim and collect trial-wise sequences for training the model. Subjects were excluded if they were left-handed, had poor scalp coupling index, discontinued participation, or had corrupted or incomplete data. Additionally, trials with acquisition errors, and trials that had data corruption during trimming were excluded from analysis. Input normalization was performed channel-wise using max-min scaling to the [1, 11] range, with 0 reserved for padding masks. Normalization bounds were set using the 98% confidence interval of a Johnson's bounded distribution, which was selected based on maximum likelihood estimates among candidate distributions in the SciPy library.

The cricothyrotomy dataset used a different optode montage, the schematic of which is included in the Supplementary Materials (Fig. S3), along with the original montage from Subedi et al. (2025). Hierarchical Task Analysis (HTA) was used to segment each procedure into qualitatively distinct subtasks. For FLS Suturing, 13 subtasks were defined per Kamat et al. (2024). ETI originally comprised six subtasks per Taylor et al. (2022), but Subtask 1 (mannequin alignment) and Subtask 6 (post-intubation checks) were excluded due to irrelevance in this protocol, resulting in four analyzed subtasks. To ensure consistent temporal analysis across trials of varying duration, temporal attention values were averaged within each defined subtask.

### 5.2. Model Development and Validation Methods:

The foundational model architecture is based on a two-layer transformer encoder [Vaswani et al., 2017]. The model processes raw fNIRS time-series data, where each trial is represented as a sequence of measurements across multiple fNIRS channels. In the first, unsupervised pretraining phase, this encoder block is trained using a masked modelling scheme adapted from the Bidirectional Encoder Representations from Transformers (BERT) model [Devlin et al., 2019], modified appropriately for time-series data. For instance, instead of randomly masking tokens, fNIRS time-series segments of lengths 3 to 11 are randomly chosen and masked across the full sequence. These segments are reconstructed using a mean squared error loss, which is backpropagated to update encoder weights. To handle variable-length fNIRS trials, sequences within each batch are padded with zeros to match the maximum sequence length present in that batch. This zero-padding is subsequently ignored during model computation, along with the masked segments, through the use of an attention padding mask, generated based on these zero-valued time steps. Pretraining is conducted over 1000 epochs with a batch size of 20 on a GPU with 4GB VRAM.

In the second, supervised phase, the pretrained encoder weights are frozen. Outputs from each encoder layer are passed to the decoder via a cross-attention mechanism. The decoder itself receives a fixed-size input tensor where each longitudinal element (or "token") corresponds to a distinct, predefined classification label category (e.g., 'task proficiency'). For this study, a single token representing 'task proficiency' is utilized. Using its cross-attention module, the decoder treats the embeddings of these label category tokens as queries to selectively attend to relevant temporal information within the encoder's processed representation of the entire fNIRS trial. This mechanism allows the decoder to generate a specific prediction for the queried label category. The architecture is designed to fully accommodate additional annotation categories (e.g. self-reported cognitive load, training history) without interfering with performance, as each category is processed independently through the decoder. The final prediction output of the decoder model is sigmoid-activated to yield binary logits for the 'task proficiency' label, indicating a pass or fail outcome. Both encoder and decoder blocks have an embedding dimension of 80 and 5 attention heads.

The temporal attention within the decoder's cross-attention layer operates by computing dot-product attention between decoder query embeddings and encoder time steps, thereby identifying salient time points for classification. A novel addition to the decoder model is the channel attention mechanism, which learns an attention map in the channel axis to identify factors to scale up signals from a subset of channels and attenuate the rest. This mechanism uses 16 attention heads with a dropout rate of 0.5 for improved generalizability and consistency, as evidenced by repeated channel selection across model initializations (see Supplementary Section). The module learns a query embedding from fixed positional encodings and applies dot-product attention across the channel (key) dimension. A residual connection adds the reweighted signals back to the original input, facilitating gradient flow since this module lies at the base of the model pipeline.

For repeatability, nine different initializations of encoder blocks were trained for masked modelling phase, and decoder classifiers were trained for each of them. Although the average variance of the output logits is significantly low (<0.03), temporal heatmaps vary with initializations. This highlights the sensitivity of temporal attention to initial conditions, and the importance of averaging the results across multiple initializations, especially if interpretation of temporal attention is of high interest.

During the few-shot adaptation phase, a light-weight two-layer dense network is trained, while all encoder and decoder weights remain frozen. Biases in the adapter are fixed (non-trainable), and adaptation is solely based on learning feature projections from new data. For each k-shot condition, 24 random training samples were selected and evaluated using a 3-fold cross-validation protocol. Final performance metrics are reported as averages over 72 evaluations.

**Data availability.**

Data used in this study will be released in future for public use after publication of all results under research currently.

**Code availability.**

The codes used in this analysis are available in a private repository in github at https://github.com/axiom5/LargefNIRS, and will be made available upon reasonable request.

Acknowledgements.

The authors gratefully acknowledge the support of this work through the Medical Technology Enterprise Consortium (MTEC) award #W81XWH2090019, and the U.S. Army Futures Command, Combat Capabilities Development Command Soldier Center STTC cooperative research agreement #W912CG2120001. The funders played no role in the study design, data collection, analysis and interpretation of data, or the writing of this manuscript.

Competing interests

All authors declare no financial or non-financial competing interests.